\documentclass[article,twocolumn,floatfix,longbibliography]{revtex4-2}
\usepackage{graphicx} 
\usepackage[caption=false]{subfig}
\usepackage{hyperref}
\usepackage{xcolor}
\usepackage{multirow}
\usepackage{lineno} 
\begin{document}
\title{ANAIS--112 three years data: a sensitive model independent negative test of the DAMA/LIBRA dark matter signal}
\author{Iv{\'a}n Coarasa\textsuperscript{1,2}}
\email{icoarasa@unizar.es}
\author{Julio Amar{\'e}\textsuperscript{1,2}}
\author{Jaime Apilluelo\textsuperscript{1,2}}
\author{Susana Cebri{\'an}\textsuperscript{1,2}}
\author{David Cintas\textsuperscript{1,2}}
\author{Eduardo Garc\'{\i}a\textsuperscript{1,2}}
\author{Mar\'{\i}a Mart\'{\i}nez\textsuperscript{1,2}}
\email{mariam@unizar.es}
\author{Miguel {\'A}ngel Oliv{\'a}n\textsuperscript{1,2,3}}
\author{Ysrael Ortigoza\textsuperscript{1,2,4}}
\author{Alfonso Ortiz~de~Sol{\'o}rzano\textsuperscript{1,2}}
\author{Tamara Pardo\textsuperscript{1,2}}
\author{Jorge Puimed{\'o}n\textsuperscript{1,2}}
\author{Ana Salinas\textsuperscript{1,2}}
\author{Mar\'{\i}a Luisa Sarsa\textsuperscript{1,2}}
\author{Patricia Villar\textsuperscript{1}}
\affiliation{\textsuperscript{1}Centro de Astropart\'{\i}culas y F\'{\i}sica de Altas Energ\'{\i}as (CAPA), Universidad de Zaragoza, Pedro Cerbuna 12, 50009 Zaragoza, Spain}
\affiliation{\textsuperscript{2}Laboratorio Subterr\'aneo de Canfranc, Paseo de los Ayerbe s.n., 22880 Canfranc Estaci\'on, Huesca, Spain}
\affiliation{\textsuperscript{3}Fundaci\'on CIRCE, Av. de Ranillas 3D, 50018 Zaragoza, Spain}
\affiliation{\textsuperscript{4}Escuela Universitaria Polit\'ecnica de La Almunia de Do\~{n}a Godina (EUPLA), Universidad de Zaragoza, Calle Mayor 5, La Almunia de Do\~{n}a Godina, 50100 Zaragoza, Spain}

\newcommand{\DL}{DAMA\slash LIBRA\ }
\newcommand{\ANAIS}{\mbox{ANAIS--112}\ }
\newcommand{\COSINE}{COSINE--100\ }

\newcommand{\ckkd}{c/keV/kg/d}
\newcommand{\keV}{keV$_{\textnormal{ee}}$}

\newcommand{\tritium}{$^{3}$H\ }
\newcommand{\Na}{$^{22}$Na\ }
\newcommand{\K}{$^{40}$K\ }
\newcommand{\NaK}{$^{22}$Na\slash$^{40}$K\ }
\newcommand{\Cd}{$^{109}$Cd\ }
\newcommand{\Cf}{$^{252}$Cf\ }
\newcommand{\I}{$^{127}$I\ }
\newcommand{\Pb}{$^{210}$Pb\ }

\newcommand{\Sen}{\mathcal{S}}

\renewcommand\labelenumi{(\theenumi)}

\begin{abstract}
Weakly interacting massive particles (WIMPs) are well-motivated candidates for dark matter.
One signature of galactic WIMPs is the annual modulation expected in a detector's interaction rate, which arises from Earth's revolution around the Sun.
Over two decades, the \DL experiment has observed such modulation with 250 kg of NaI(Tl) scintillators, in accordance with WIMP expectations but inconsistent with the negative results of other experiments.
The signal depends on the target material, so to validate or refute the DAMA result, the experiment must be replicated using the same material. This is the goal of the \ANAIS experiment, currently underway since August 2017 with 112.5~kg of NaI(Tl).
In this work, we present a reanalysis of three years of data employing an improved analysis chain to enhance the experimental sensitivity. The results presented here are consistent with the absence of modulation and inconsistent with DAMA's observation at nearly 3$\sigma$ confidence level, with the potential to reach a 5$\sigma$ level within 8 years from the beginning of the data collection.
Additionally, we explore the impact of different scintillation quenching factors in the comparison between \ANAIS and DAMA\slash LIBRA.
\end{abstract}
\maketitle
\section*{Introduction}
\label{sec:intro}

The question of dark matter's nature remains still unanswered almost a century after its initial proposal~\cite{Zwicky:1933gu,Workman:2022ynf}.
Among the myriad candidates proposed to constitute the dark matter (DM), Weakly Interacting Massive Particles (WIMPs) have attracted significant attention due to both their compelling theoretical motivation and potential detectability~\cite{Bertone:2004pz}.
\\
WIMPs, hypothesized to interact weakly with ordinary matter, would have been abundantly produced in the early universe, offering a natural explanation for the observed abundance of DM~\cite{Planck2018}. Furthermore, theories beyond the Standard Model of particle physics, such as Supersymmetry or extra dimensions, present potential candidates for WIMPs.
Efforts to detect WIMPs have followed different strategies, including searching for their annihilation products 
as an excess in the fluxes of cosmic messengers reaching the Earth~\cite{Conrad:2017pms}, 
or identifying new physics phenomena in colliders, such as signals of new mediators or events with missing energy resulting from dark matter production~\cite{Boveia:2018yeb}. 
Additionally, attempts to detect WIMPs directly have been performed by searching for their interaction 
with sensitive detectors on Earth~\cite{Goodman:1984dc},
which predominantly involves elastic scattering off atomic nuclei.
While significant portions of the parameter space for benchmark (generic) particle candidates have been excluded using these 
methods~\cite{Billard:2021uyg},
our limited knowledge of the underlying models makes these results strongly model-dependent. Furthermore, distinguishing backgrounds from the signal is challenging.
Hence, it is crucial to identify a characteristic signature of dark matter.
Among the few proposed ones, annual modulation stands out as particularly compelling~\cite{Drukier:1986tm, Freese:1987wu}. 
\\
The flux of WIMP particles on Earth depends on the relative velocity between Earth and the DM halo. 
As Earth, along with the Solar System, moves towards the Cygnus constellation during its orbit 
around the galactic center, the Earth's revolution around the Sun introduces a minor correction to its velocity relative to the halo.
The differential scattering rate $R$ as a function of the nuclear recoil energy $E_{\text{NR}}$ and the time is~\cite{Lewin:1995rx}
\begin{equation}
\frac{dR(E_{\text{NR}},t)}{dE_{\text{NR}}} = \frac{N_T \rho_0}{m_\chi}\int_{v_{\text{min}}}^{v_{\text{max}}} v f(\vec{v},t)
\frac{d\sigma(E_{\text{NR}}, v)}{dE_{\text{NR}}}d^3\vec{v} , 
\end{equation}
where $N_T$ is the number of target nuclei, $\rho_0$ is the local DM density, $m_\chi$ is the mass of the DM particle,
$\vec{v}$ is the DM velocity in the detector's rest frame
and $v$ is its modulus,
$f$ is the velocity distribution of DM particles in the detector's rest frame,
$v_{\text{max}}$ is the maximum velocity of DM particles in the detector's rest frame corresponding to the escape speed of the Milky Way,
$v_{\text{min}}=\sqrt{m_{\text{N}}E_{\text{NR}}/2\mu_{\text{N}\chi}^2}$ 
is the minimum velocity of DM particles that can produce a nuclear recoil of energy $E_{\text{NR}}$ off a nucleus with mass 
$m_{\text{N}}$, where $\mu_{\text{N}\chi}$ is the reduced mass of the WIMP-nucleus system,
and $\sigma$ is the WIMP-nucleus scattering cross section.
The maximum recoil energy $E_{\text{NR}}^{\text{max}}=2\mu_{\text{N}\chi}^2v^2/m_{\text{N}}$ for typical WIMP velocities 
O(200\,km\,s$^{-1}$) ranges from approximately 10 to 100\,keV, depending on $m_{\text{N}}$.
\\
The velocity distribution function $f$ can be calculated from the velocity distribution function in the
Galactic reference system $f_{\text{gal}}$ through a Galilean transformation 
$f(\vec{v},t)=f_{\text{gal}}(\vec{v}+\vec{v}_{\text{E}}(t))$, where
$\vec{v}_{\text{E}}$ is the Earth's velocity in the Galactic rest frame.
$f_{\text{gal}}$ is truncated at the Milky Way escape speed~\cite{Baxter:2021pqo}, $v_{\text{esc}}$=544\,km\,s$^{-1}$. $\vec{v}_{\text{E}}$ comprises three primary components~\cite{Baxter:2021pqo}:
(1) the motion of the local standard of rest, that in 
galactic coordinates is given by $(0,v_0,0)$, with $v_0$=238\,km\,s$^{-1}$, (2) the Sun's peculiar motion, which is $(11.1, 12.2, 7.3)$\,km\,s$^{-1}$ and (3)
the orbital motion around the Sun, which can be well approximated by a circular orbit tilted $\theta\approx60^{\circ}$ 
with respect to the Galactic plane, at an orbital speed of $v_{\text{orb}}=$29.8\,km\,s$^{-1}$.
A reasonably accurate approximation for the Earth's speed is given by
\begin{equation}
v_{\text{E}}=v_\odot+v_{\text{orb}}\cos\theta \cos(\omega(t-t_0)),
\end{equation}
where $v_\odot$=$v_0$+12.2\,km\,s$^{-1}$, $\omega=\frac{2\pi}{365}$\,d$^{-1}$ and the phase $t_0$ depends on the specific halo model considered, 
but in most virialized models is about June 2, when the combined velocities reach their maximum.
\\
Hence, there are slightly more WIMPs with high speeds in the detector's rest frame during the summer, 
and conversely, more WIMPs with low speeds during the winter. 
This leads to a modulation in the differential rate, with the highest rate occurring in summer 
for larger nuclear recoil energies and in winter for smaller ones. 
Considering that the variation in Earth’s speed between summer and winter amounts to roughly 6\% of the average velocity, 
the differential rate can be approximated using a Taylor series
\begin{equation}
\frac{dR}{dE_{\text{NR}}} \approx 
\left(\frac{dR}{dE_{\text{NR}}}\right)_{v_{\text{E}}=v_\odot}+\Delta(E_{\text{NR}})\cos(\omega(t-t_0)),
\end{equation}
with 
\begin{equation}
\Delta(E_{\text{NR}})=\left(\frac{d^2R}{dE_{\text{NR}}dv_{\text{E}}}\right)_{v_{\text{E}}=v_\odot}v_{\text{orb}}\cos\theta .
\end{equation}
Therefore, the expected signal of dark matter 
integrated over a certain energy window (here denoted by $k$)
 can be expressed as the sum of a constant term plus 
a term modulated with an annual period:
\begin{equation}
R_k(t) \approx R_{0,k} + S_{m,k} \cos(\omega (t-t_0)).
\end{equation}
\\
If the experimental threshold is low enough, the sign change in $S_{m,k}$ should be observed at a characteristic energy dependent on the target nucleus and WIMP masses. The observation of this phase shift would allow to determine the WIMP mass. The annually modulated signal is faint, corresponding only to a small percentage of the total signal.
However, the requirements it must meet to be interpreted as produced by WIMPs in the galactic halo are very restrictive:
it must have the correct amplitude, phase, and period, and occur only in the low-energy region. 
\\
For over 20 years, the \DL experiment has observed a modulation in its data that satisfies these criteria, 
thus representing a  strong indication of dark matter detection~\cite{Bernabei:2020mon}. 
The DAMA/NaI experiment began in 1995 at the Laboratori Nazionali del Gran Sasso, Italy, 
with 100\,kg of NaI(Tl) scintillators and an energy threshold set at 2\,keV~\cite{Bernabei:1996vj}. 
After 7 years, the experiment upgraded to the LIBRA setup, scaling up the detector mass to 250 kg
(DAMA/LIBRA-phase1)~\cite{DAMA:2008jlt}. 
Subsequently, after 7 additional years of data collection, 
all photomultiplier tubes were replaced with others with enhanced quantum efficiency, 
thus reducing the energy threshold to 1\,keV (DAMA/LIBRA-phase2)~\cite{Bernabei:2018jrt}.
The experiment
is still ongoing, with an exposure that has already reached 2.86\,ton$\times$yr over 22 independent annual cycles~\cite{Bernabei:2021kdo}.
The modulation signal observed by DAMA is relatively large in amplitude, 
$S_m^{\text{DAMA}}$=10.5$\pm$1.1 (10.2$\pm$0.8)\,counts\,keV$^{-1}$\,ton$^{-1}$\,d$^{-1}$ 
for [1--6] ([2--6])\,keV energy region~\cite{Bernabei:2020mon}.
That corresponds to nucleon cross sections 
of the order of $10^{-40}$--$10^{-41}$\,cm$^2$ when interpreted as a WIMP with spin- and isospin-independent coupling. Such a signal should have already been observed by other 
direct detection experiments, which, however, do not observe events above their estimated backgrounds
and can exclude the \DL signal with a very high confidence level~\cite{XENON:2023cxc,DarkSide-50:2022qzh,PandaX-4T:2021bab,LZ:2022lsv,SuperCDMS:2018gro,EDELWEISS:2019vjv,CRESST:2019jnq,CRESST:2022jig,PICO:2019vsc}.
Nevertheless, the comparison between experiments strongly depends on the model employed for the WIMP and its velocity distribution in the galactic halo. Additionally, the lack of alternative explanation to date for the \DL signal makes it imperative to seek independent confirmation using the same target material.
\\
This is the goal of several experiments, either completed (DM-ICE~\cite{DM-Ice:2016snk}, COSINE--100~\cite{COSINE-100:2021zqh}), in data taking (ANAIS--112~\cite{Amare:2018sxx}),
under construction (COSINE--200~\cite{COSINE:2020egt}, COSINUS~\cite{cosinus2016}) or in R\&D phase (SABRE~\cite{Zani:2022ysd}, PICOLON~\cite{picolon2021}, ASTAROTH~\cite{DAngelo:2022kse}, ANAIS+~\cite{Amare:2022ncr}).
To accurately verify the DAMA signal, an experiment must  possess the capability to replicate it
with high statistical significance (which requires ultra-low background levels with a threshold energy at or below 1\,keV, 
large exposure and operational stability) besides a thorough understanding of the detector's response function~\cite{Sarsa_2024}.
This implies addressing various factors including the non-linear energy response, energy resolution, and the efficiencies for triggering and event acceptance. Moreover, because the conversion into light of the energy released by highly ionizing particles is quenched compared to electrons, 
it is particularly important to 
consider the conversion factor between the energy deposited 
by a nuclear recoil ($E_{\text{Na}},E_{\text{I}}$) 
and the energy estimated through a calibration performed with beta/gamma sources,
or electron-equivalent energy ($E_{\text{ee}}$).
As of today, the scintillation quenching factors $Q_{\text{Na,I}}=E_{\text{ee}}/E_{\text{Na,I}}$
cannot be calculated but must be measured and they have been shown to depend on energy.
The available measurements~\cite{Cintas:2024pdu,Bignell:2021bjx, Joo:2017aws, Stiegler:2017kjw,Xu:2015wha,Collar:2013gu,Chagani:2008in,Gerbier:1998dm,Tovey:1998ex,Bernabei:1996vj,
Spooner:1994ca} span in the range 0.10--0.30 for $Q_{\text{Na}}$ and 0.05--0.09 for $Q_{\text{I}}$
for nuclear recoil energies below 100\,keV.
\\
Under the assumption that the DAMA signal is originated by WIMPs interacting with nuclei, 
in order to compare the signals from different experiments, it is necessary to assume that the quenching factors 
$Q_{\text{Na,I}}$ are the same for different NaI(Tl) detectors, or alternatively, 
measure the quenching factor in each case and correct the energy scale.
In this paper, all the energies in relation to 
NaI(Tl) detectors are given as electron-equivalent energies.
As it will be explained in the last section of this article, differences in quenching factors are the only systematic effect that could compromise a direct comparison of experiments using the same target. In addition to improve the understanding of the scintillation quenching factors in NaI(Tl), this systematic effect could be handled by strongly reducing the energy threshold well below that of DAMA/LIBRA. ANAIS+ is one of the R\&D projects pursuing this goal, by replacing the PMTs by SiPMs and operating the detector at 100~K.
\\
The \ANAIS experiment is composed of 112.5\,kg of NaI(Tl) distributed among nine scintillator units of 12.5\,kg each, constructed by Alpha Spectra, Inc., Colorado, US. Each crystal is coupled to two photomultiplier tubes (PMTs) and each module is triggered by the coincidence between the two PMT signals within a 200~ns window. \ANAIS has been collecting data at the Canfranc Underground Laboratory, Spain, 
since August 3, 2017. 
The detectors are shielded by 10~cm of
archaeological lead, 20~cm of low activity lead, an anti-radon
box (continuously flushed with radon-free nitrogen
gas), a  muon veto made up of 16 plastic
scintillators, and 40~cm of polyethylene bricks and water tanks acting as neutron moderator.
\\
Previous \ANAIS publications have provided detailed descriptions of the experimental setup,
data acquisition system (DAQ)
and detector performance after the first year~\cite{Amare:2018sxx}, 
the background model~\cite{Amare:2018ndh}, sensitivity prospects~\cite{Coarasa:2018qzs}, 
preliminary annual modulation results for 1.5 and 2~years~\cite{Amare:2019jul, Amare:2019ncj}, 
results for 3~years of exposure~\cite{Amare:2021yyu}, and the development of a machine-learning-based 
analysis protocol for filtering non-bulk scintillation events~\cite{Coarasa:2022zak}. 
\\
In this work, we present a reanalysis of the data collected over the first 3 years, totaling an exposure of 312.53 kg$\times$yr. 
For the event selection, we use the analytical tools outlined in \cite{Coarasa:2022zak}. Additionally, we have implemented several enhancements 
to the \ANAIS analysis pipeline, including improvements in energy calibration, quality cuts, and efficiency calculation, 
which are thoroughly described in the Methods section.
\\
To search for a modulation we perform a chi-squared minimization on the event counts observed by the 9 \ANAIS detectors over time, in the two energy regions studied by DAMA ([1--6] and [2--6]\,keV). 
The results are compatible with the null hypothesis and incompatible
with the \DL signal at 3.7$\sigma$ and 2.6$\sigma$ confidence level (C.L.)
for the [1--6] and [2--6]\,keV energy regions, respectively.
Furthermore, the sensitivity of the experiment has improved as anticipated in Ref.~\cite{Coarasa:2022zak}, confirming our expectations of achieving 
5$\sigma$ sensitivity by 2025.
\\
We also investigate how the comparison between \ANAIS and \DL is influenced under the hypothesis that the quenching factors 
for sodium and iodine
recoils are different in both detectors, as recent dedicated measurements suggest~\cite{Cintas:2024pdu}. 
Considering energy-independent quenching factors, the results of ANAIS are incompatible with the DAMA signal at 3$\sigma$ C.L.
\section*{Results and Discussion}
\label{sec:res}
\subsection*{\ANAIS experimental performance and exposure}

\begin{figure}
    \centering
    \includegraphics[width=0.48\textwidth]{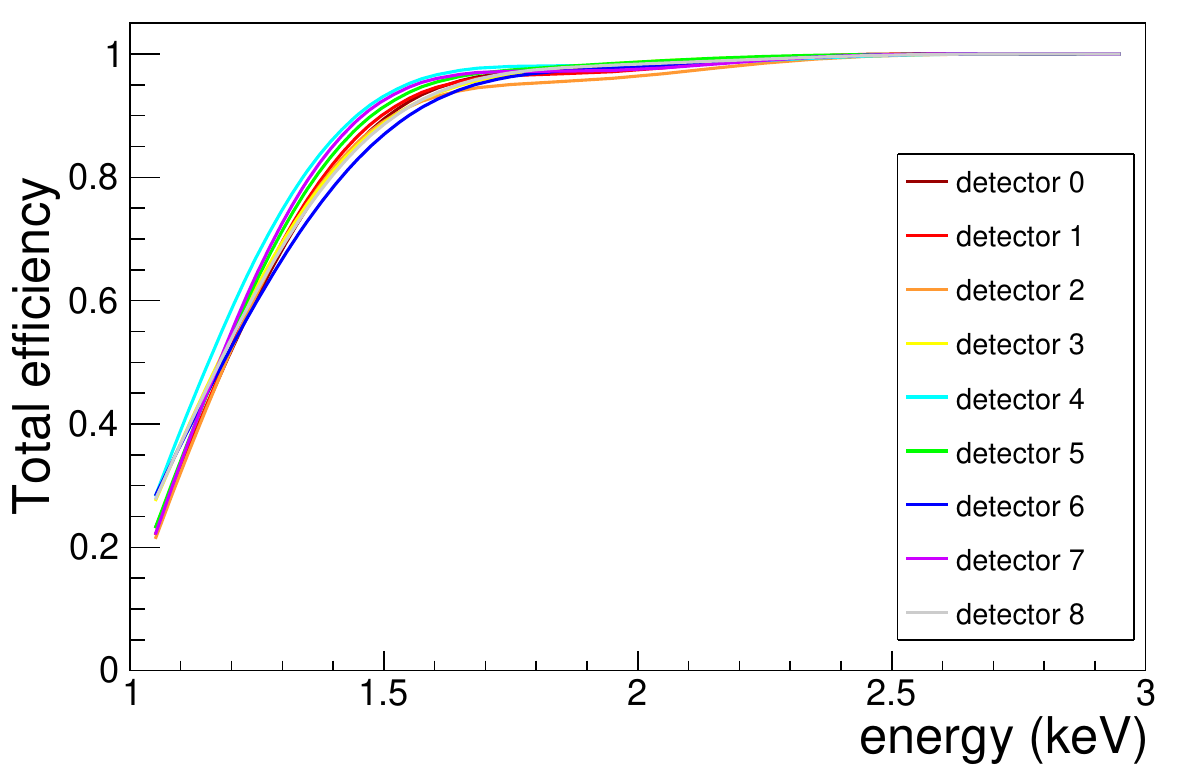} 
    \caption{\textbf{Total detection efficiency in all the \ANAIS modules as a function of energy.} It has been obtained by combining the trigger efficiency and the BDT cut efficiency (see the Methods Section).}
    \label{fig:eff}
\end{figure}

The \ANAIS detection rate remained fairly stable at about 5\,Hz, 
dominated by fast Cherenkov light produced in the photomultipliers and random coincidences of 
dark current events in the two PMTs of each module within 
the coincidence window.
Light collection stayed consistently high and homogeneous throughout the nine modules, averaging
14.6\,photoelectrons (phe)\,keV$^{-1}$ during the first year (with a 
standard deviation of 0.8\,phe\,keV$^{-1}$) and 14.4\,phe\,keV$^{-1}$ (0.7\,phe\,keV$^{-1}$ standard deviation) during the third year. 
The gain of the PMTs was stable at the 3\% level for all modules, except for D4 and D5, for which we changed the 
high voltage bias after the first year of operation. 
These drifts were monitorized and corrected with the periodic $^{109}$Cd calibrations.
\\
The trigger efficiency is 100\% down to 1.5\,keV and remains above 95\% at 1\,keV~\cite{Amare:2018sxx}. 
However, the analysis threshold is set to 1~keV because of the decrease of the acceptance efficiency for the selection of bulk scintillation events down to 20\%--30\%, depending on the detector (see Fig.~\ref{fig:eff} and the Methods Section).
%
\\
Table~\ref{tab:exposure} summarizes the accumulated exposure for the three years
of data analyzed in this work, calculated as the product of the
total mass times the live time. It also details
the dead time (measured using latched counters during the data-taking), down time (primarily due to bi-weekly $^{109}$Cd calibrations), 
percentage of periods rejected in the analysis, and the corresponding effective exposure after subtracting them. In addition to the criteria outlined in the Methods section 
for the rejection of high-rate periods, we remove events arriving within 1\,s from a muon interaction in the veto.
Scintillation time constants as long as 300 ms have been observed for high-energy $\mu$ events in NaI(Tl)~\cite{Cuesta:2013vpa}, therefore this criterion helps to prevent the DAQ from triggering numerous low-energy false events after a muon's passage through the scintillator. Additionally, it also rejects potential secondary particles generated by a cosmic muon in the detector or its shielding.
\\
\setlength{\tabcolsep}{1em}
\begin{table*}[htbp]
\centering
\begin{tabular}{ccccccc}
\hline\hline
Time period & Exposure & Dead time & Down time & \multicolumn{2}{c}{Rejected periods (\%)} & Effective exposure  \\
 & (kg$\times$yr) & (\%) & (\%) & muon cut & rate cut & (kg$\times$yr)\\
\hline
Aug 3, 2017 – July 31, 2018 & 104.80 & 2.88 & 3.20 & 2.64 & 0.57 & 101.22\\
Aug 1, 2018 – Aug 28, 2019 & 115.39 & 2.07 & 2.42 & 2.64 & 0.48 & 111.63\\
Aug 29, 2019 – Aug 13, 2020 & 102.86 & 2.38 & 2.54 & 2.53 & 0.42 &  99.68\\
\hline\hline
\end{tabular}
\caption{\label{tab:exposure} 
\textbf{Summary of the accumulated exposure for the three years of data analyzed in this work.}
For each of the three years of data collection, first column: start and end dates; second column: exposure calculated by multiplying live time by mass; third column: percentage of dead time; fourth column: percentage of down time; fifth and sixth columns: percentages of time with respect to the live time corresponding to the two types of rejected periods (one second after a muon passage and trigger cut, respectively); last column: effective exposure after subtracting the rejected periods.
}
\end{table*}
\subsection*{Energy spectrum at low energy and background modelling}
\begin{figure*}
    \centering
    \includegraphics[width=\textwidth]
    {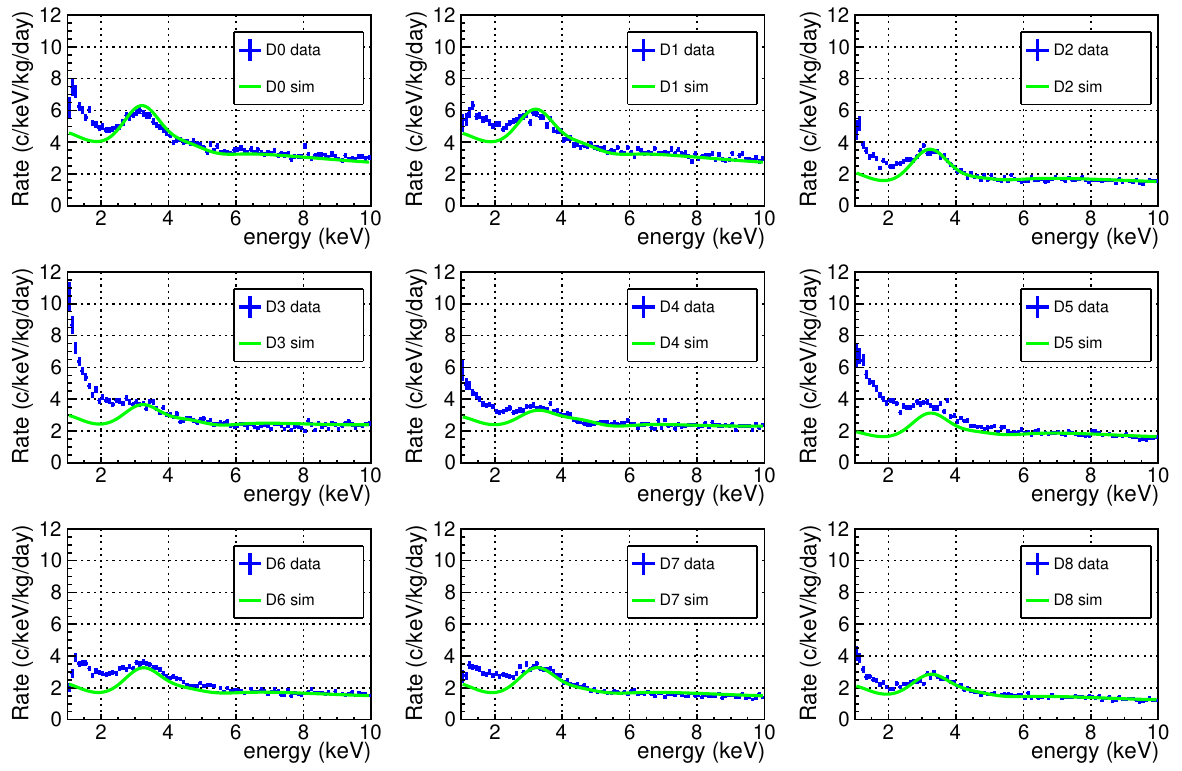} 
    \caption{\textbf{\ANAIS detectors' low energy background spectra and Monte Carlo background model.} Each panel corresponds to one of the nine detectors of \ANAIS (see the Introduction section), labeled D0 through D8. Blue points: single-hit energy spectrum measured in the ROI for each detector after event selection and efficiency correction. 
    Data correspond to the $\sim$10\% data unblinded for the first three years of operation.
    Green line: Monte Carlo background model following~\cite{Amare:2018ndh}.
}
    \label{fig:bkg3x3}
\end{figure*}

In our DM analysis, we focus on events where the energy deposition occurs in only one of the nine detectors, which are referred to in the following as single-hit events.
Those falling within the region of interest (ROI), from [1--6]\,keV, 
are blinded during the tuning of the analysis procedure. Only $\sim$10\% of the data is unblinded to evaluate the background and assess the experimental performance.
\\
The resulting low-energy spectrum for each detector is presented in Fig.~\ref{fig:bkg3x3}, based on the $\sim$10\% of data unblinded.
It has been corrected by the total detection efficiency, calculated as the product of the trigger and event selection efficiencies. 
\\
In the figure, we also show our Monte Carlo (MC) background model (a comprehensive review can be found in \cite{Amare:2018ndh}).
The model takes as input independent estimates of the different radioactive contaminations.
The background in the ROI is dominated by radioactive contamination of the NaI(Tl) crystal, 
particularly by $^{210}$Pb out of equilibrium. This isotope is found in varying quantities in the different crystals, both in the bulk and in the surface, 
being higher for the earliest detectors constructed, D0 and D1, at a level of 3.15\,mBq\,kg$^{-1}$. 
After introducing improvements in the growth and purification process, 
the $^{210}$Pb level in detectors D2 to D8 decreased to values in the range of 0.7--1.8\,mBq\,kg$^{-1}$. 
Another common contaminant of NaI(Tl) is $^{40}$K, due to its chemical affinity. 
In the ANAIS crystals, it is present at levels of around 1\,mBq\,kg$^{-1}$ and is responsible of the peak at 3.2\,keV
visible in the spectra of Fig.~\ref{fig:bkg3x3}. 
This energy is released by the de-excitation of the atomic K-shell following 
an electron capture (EC), when the high energy $\gamma$ ray escapes. Sometimes, this $\gamma$
can hit another crystal, producing a coincident event that allows us to tag the low energy 
deposition and use it to both estimate the amount of $^{40}$K in the crystal and select these low-energy events
for calibration and efficiency calculation.
The cosmogenic isotope $^{22}$Na is also present in the crystals and 
produces a background in the ROI of similar origin, but in this case the K-shell relaxation energy is at 0.87\,keV,
and is only marginally present in the region between 1 and 2\,keV. 
$^3$H is another cosmogenic isotope that contributes significantly to the ROI. It is a pure beta emitter with 
an end-point at 18.591\,keV. Other cosmogenic isotopes of tellurium and iodine contribute to the region of interest through 
L- and K-shell EC emissions~\cite{Amare:2014bea}. Their half-lives are shorter (between $\sim$10 and $\sim$150 days) and are not relevant to the total background of the experiment, but they are important for explaining the evolution of the background, especially in the latest detectors arriving at Canfranc (D6, D7, and D8).
The agreement between the data and the model is very good down to 3\,keV. Below this energy, 
there appears to be a component that is not well explained by the model. 
It could be optical noise that escapes the event selection or 
some radioactive background contribution missed in the model.
Present studies with a larger digitization window (8 microseconds instead of 1.2) point to the first explanation as the main cause: this energy region seems to be dominated by a population of events with a time scale not compatible with NaI(Tl) scintillation and asymmetric in the light sharing between both PMTs. Light emissions at the PMTs (scintillation, corona discharges, etc.) could be responsible for these events. For the analysis presented in this paper, it acts as a background component whose evolution remains constant in time.
\subsection*{Annual modulation results}
\begin{figure}
    \centering
    \includegraphics[width=0.48\textwidth]{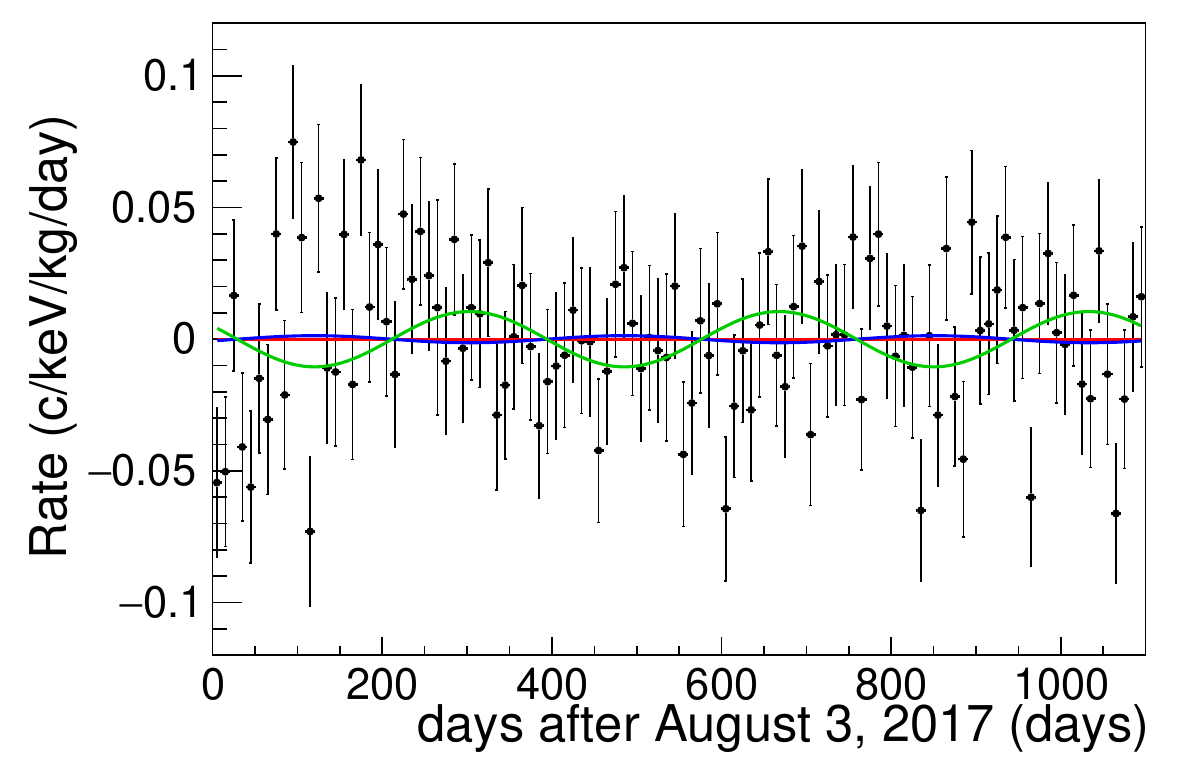} 
    \caption{
    \textbf{Fit results for the combined data of the 9 detectors in the [1--6]\,keV energy region after subtracting the background term.}
    The data points represent the combined data of the 9 detectors for the energy region [1--6]\,keV after subtracting the background term from Eq.~\ref{eq:modFit}. Error bars have been calculated by combining the standard deviation of the data from each detector. Blue and red lines are the result of the modulation and null hypothesis, respectively, after subtracting the background term from Eq.~\ref{eq:modFit}.
    The modulation observed by DAMA/LIBRA is shown in green.}
    \label{fig:residualsEvol16}
\end{figure}

To perform an independent test of the \DL signal, we look for an annual modulation in the \ANAIS data, 
but following a slightly different method as the DAMA collaboration does. DAMA calculates the residual rate of anticoincidence events 
vs time by subtracting to the total rate the annual average.
These residuals are then fitted to a function of the form $A \cos(\omega(t-t_0))$.
While it has been noted that this approach may introduce a bias in the fit for slowly varying backgrounds~\cite{Buttazzo:2020bto,Messina:2020pnt,COSINE-100:2022dvc},
this explanation seems unlikely for the DAMA signal: the phase obtained by DAMA would correspond to a slightly increasing background, 
which is challenging to explain, and no bias is observed above the energy region where the DM signal is anticipated~\cite{Bernabei:2020mon}.
\\
To avoid any potential systematic effects, we adopt a different approach, directly looking for the modulation in the overall event count over time through
a least squares fit.  We define the $\chi^2$ function as follows:
\begin{equation}
    \chi^2 = \sum_{i,d} \frac{(n_{i,d} - \mu_{i,d})^2}{\sigma_{i,d}^2} ,
\end{equation}
where 
$n_{i,d}$ represents the number of events in the ROI in the time bin $t_i$ for detector $d$, obtained by correcting 
the measured event count using the live time for that specific temporal bin and detector, along with the corresponding acceptance efficiency,
$\sigma_{i,d}$ is the Poisson uncertainty associated with the event count, also corrected by the corresponding live time and efficiency, and
$\mu_{i,d}$ denotes the expected number of events in that particular time bin and detector, including a hypothetical DM component.
\\
Given the presence of radioactive isotopes with  half-lives of the order of few years in the detectors, 
primarily $^{210}$Pb (T$_{1/2}$=22.3\,yr), $^{3}$H (T$_{1/2}$=12.3\,yr) and $^{22}$Na (T$_{1/2}$=2.6\,yr),
$\mu_{i,d}$ diminishes over time. 
For detectors D6, D7 and D8, also Te and I cosmogenic isotopes contributions are relevant.
Accurately modeling this background rate decrease is crucial to avoid biasing the fit. We employ the following expression to describe it:
\begin{equation}
    \mu_{i,d}=[R_{0,d}(1+f_d\phi_{bkg,d}^{MC}(t_i))+S_m \cos(\omega(t_i-t_0))]M_d\Delta E \Delta t,
    \label{eq:modFit}
\end{equation}
where $\phi_{bkg,d}^{MC}$ is a probability distribution function 
sampled from the MC model,
describing the background evolution at time bin $t_i$ for detector $d$,
$M_d$ is the mass of every module, and $\Delta E$ and $\Delta t$ represent energy and time intervals, respectively.
$R_{0,d}$ and $f_d$ are free parameters for each detector, 
while $S_m$ represents the DM annual modulation amplitude. 
It is set to 0 to test the null hypothesis and allowed to vary freely for the modulation hypothesis.
It is worth noting that the time-invariant component $R_{0,d}$
includes both the background produced by isotopes with 
long half-lives and components of noise not explained by the model, as well as the constant component of a hypothetical contribution from DM.
\\
In our fit, the period is fixed at one year and the phase set to June 2.
In this way, we can directly compare our results with those of the \DL 
experiment as they appear in Ref.~\cite{Bernabei:2020mon}.
We perform two independent fits, one in the [2--6]\,keV region, 
which can be compared with the results from the total accumulated exposure of DAMA/NaI and DAMA\slash LIBRA, 
and another in the [1--6] keV region, allowing us to study the results of DAMA/LIBRA-phase2.
\begin{figure*}
    \centering
    \includegraphics[width=0.85\textwidth]{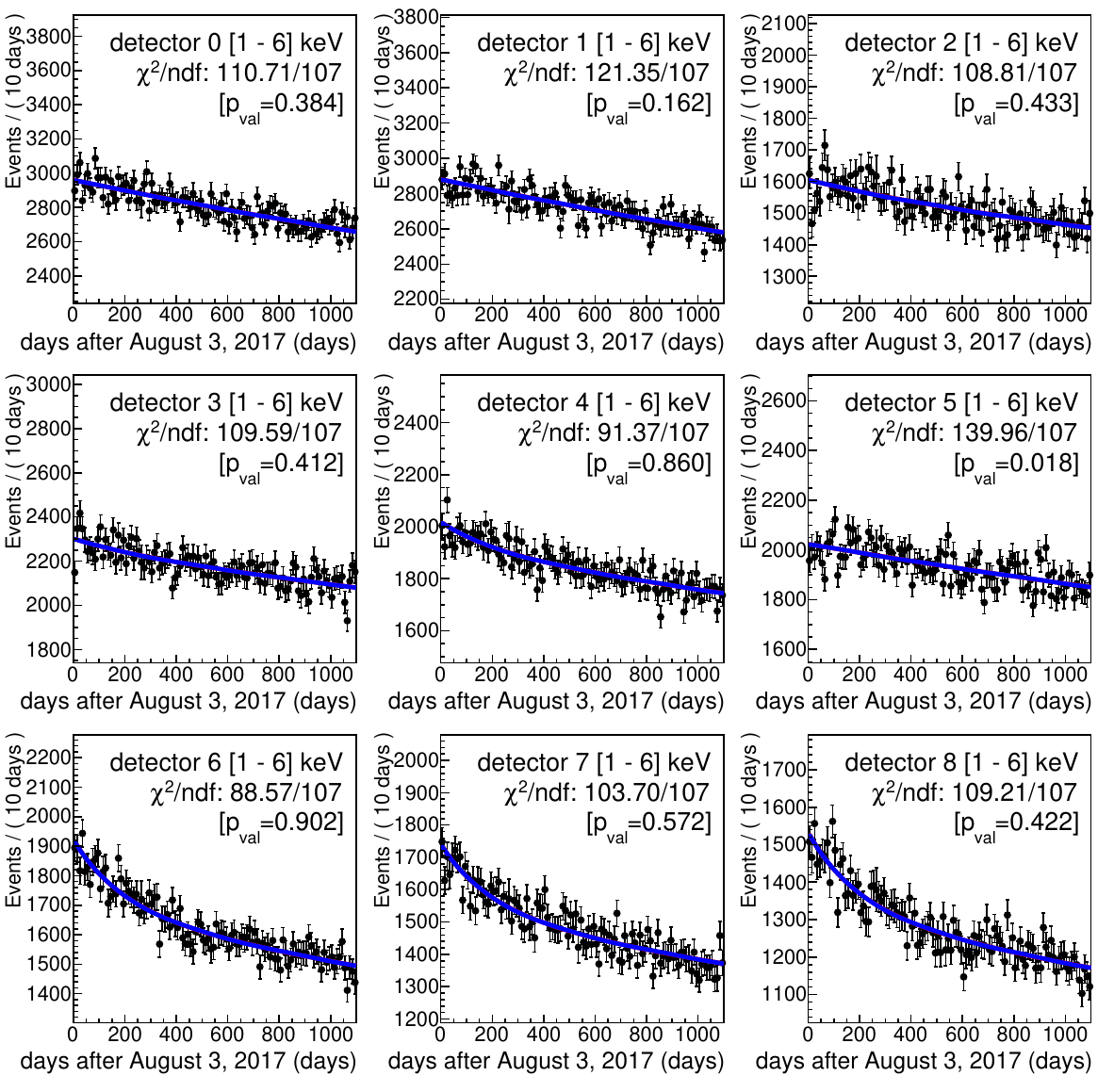} 
    \caption{
\textbf{Fit results for data from the nine \ANAIS modules in the [1--6]\,keV energy range, under both the modulation and null hypotheses}. Each panel corresponds to one of the nine detectors of ANAIS-112. The error bars on the data points represent the standard deviation of the observed rate of events combined with the efficiency uncertainty. The blue line shows the result of the modulation hypothesis fit, while the red line represents the result of the null hypothesis, although it is generally masked by the blue line and not visible. Each panel also displays the $\chi^2$ divided by the degrees of freedom (NDF) of the fit for each detector, along with the corresponding p-value. The global results of the fit are: for the null hypothesis, $\chi^2$/NDF = 982.20/972, corresponding to a p-value~=~0.403, and for the modulation hypothesis, $\chi^2$/NDF~=~982.07/971, corresponding to a p-value~=~0.395. The best-fit modulation amplitude in the latter case is $S_m = (-1.3 \pm 3.7)$ (counts keV$^{-1}$ ton$^{-1}$ day$^{-1}$).
}
    \label{fig:rateEvol16}
\end{figure*}
%
\begin{figure*}
    \centering
    \includegraphics[width=0.85\textwidth]{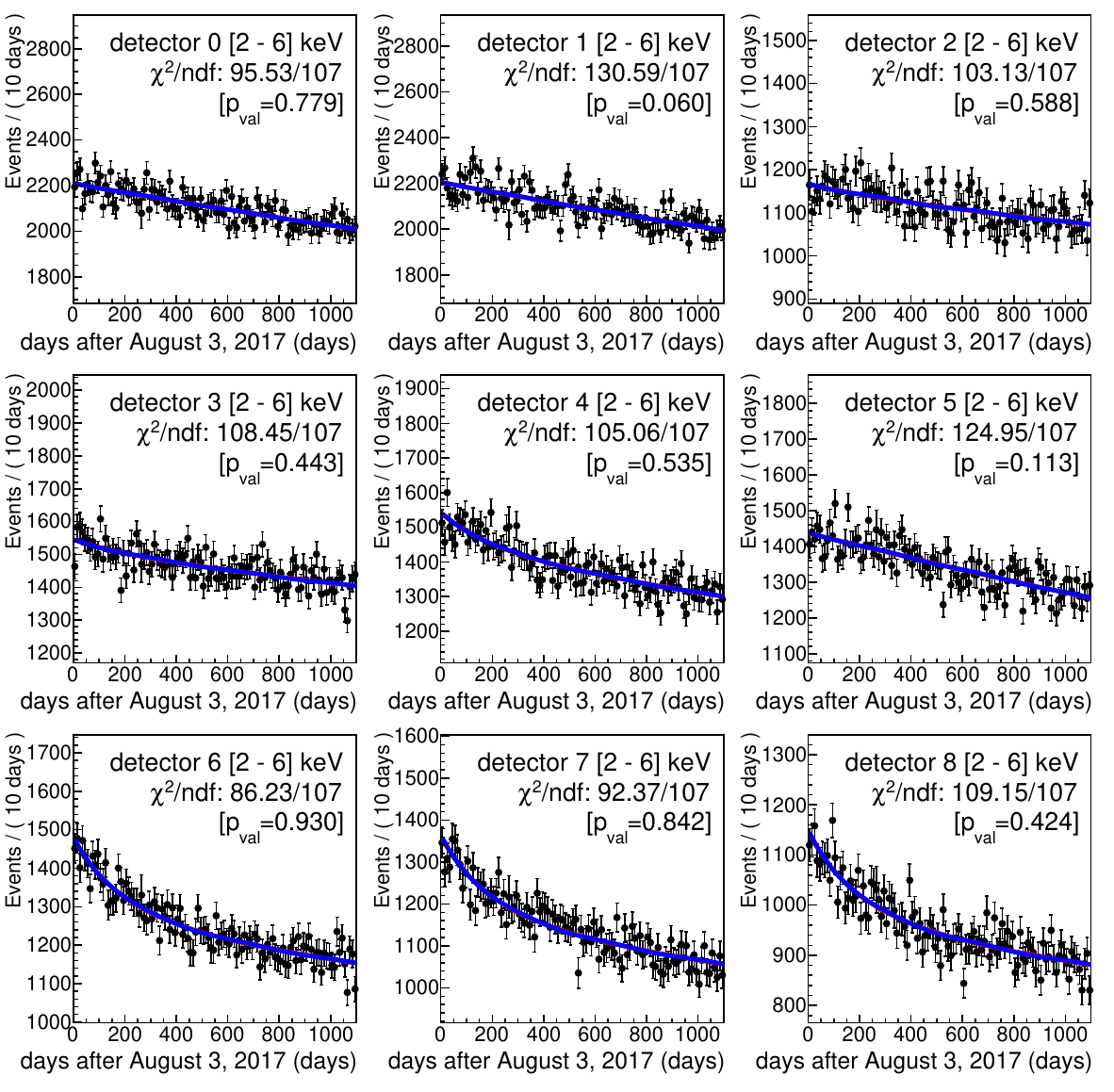} 
    \caption{
    \textbf{Fit results for data from the nine \ANAIS modules in the [2--6]\,keV energy range, under both the modulation and null hypotheses}. Each panel corresponds to one of the nine detectors of ANAIS-112. The error bars on the data points represent the standard deviation of the observed rate of events combined with the efficiency uncertainty. The blue line shows the result of the modulation hypothesis fit, while the red line represents the result of the null hypothesis, although it is generally masked by the blue line and not visible. Each panel also displays the $\chi^2$ divided by the degrees of freedom (NDF) of the fit for each detector, along with the corresponding p-value. The global results of the fit are: for the null hypothesis, $\chi^2$/NDF = 955.25/972, corresponding to a p-value~=~0.643, and for the modulation hypothesis, $\chi^2$/NDF~=~954.56/971, corresponding to a p-value~=~0.641. The best-fit modulation amplitude in the latter case is $S_m = (3.1 \pm 3.7)$ (counts keV$^{-1}$ ton$^{-1}$ day$^{-1}$).
}
    \label{fig:rateEvol26}
\end{figure*}
\\
The results of the $\chi^2$ minimization for the null and modulation hypothesis 
are displayed in Figs.~\ref{fig:rateEvol16} and \ref{fig:rateEvol26}
for data in the [1--6] and [2--6]\,keV energy regions, respectively, grouped in 10-day bins.
The possible presence of a bias in the fit was studied in Ref.~\cite{Amare:2021yyu} using a large set of Monte Carlo pseudoexperiments, sampled from the background models with no modulation or with the modulation amplitude observed by DAMA/LIBRA. In all cases, the bias was found to be compatible with zero or negligible. In the present analysis, only the experimental data have changed with respect to Ref.~\cite{Amare:2021yyu}, so the bias study remains valid in this case. The results do not exhibit dependence on the time bin size for values between 5 and 30\,days.
\\
\ANAIS results are consistent with the null hypothesis, with p-values of 0.40 and 0.64 for [1--6] and [2--6]\,keV energy regions, respectively. 
Best fits for the modulation hypothesis are consistent with the absence of modulation 
within one standard deviation
in both regions, with modulation amplitudes 
$S_m$\,=\,--1.3\,$\pm$\,3.7\,counts\,keV$^{-1}$\,ton$^{-1}$\,d$^{-1}$ and 3.1\,$\pm$\,3.7\,counts\,keV$^{-1}$\,ton$^{-1}$\,d$^{-1}$, respectively.
The $\chi^2$ divided by the number of degrees of freedom (NDF) and corresponding 
p-values are also calculated 
separately for the data of every module and displayed in the legend of each panel. 
The p-values are  greater than 0.05 in all cases, except for D5 in the [1--6]\,keV 
energy region. Notably, these values have improved compared to the 
previous analysis~\cite{Amare:2021yyu} 
for the energy region [1--6]\,keV, presumably due to improved filtering of noise events below 2\,keV.
For illustrative purposes, Fig.~\ref{fig:residualsEvol16} shows the fit results after subtracting the background term from Eq.~\ref{eq:modFit} in both the fitting functions and the data for the energy region [1--6]\,keV for the combined data of the 9 detectors. The modulation observed by DAMA/LIBRA is shown in green.
\subsection*{Experimental sensitivity and prospects}

\begin{figure}
    \centering
    \includegraphics[width=0.48\textwidth]{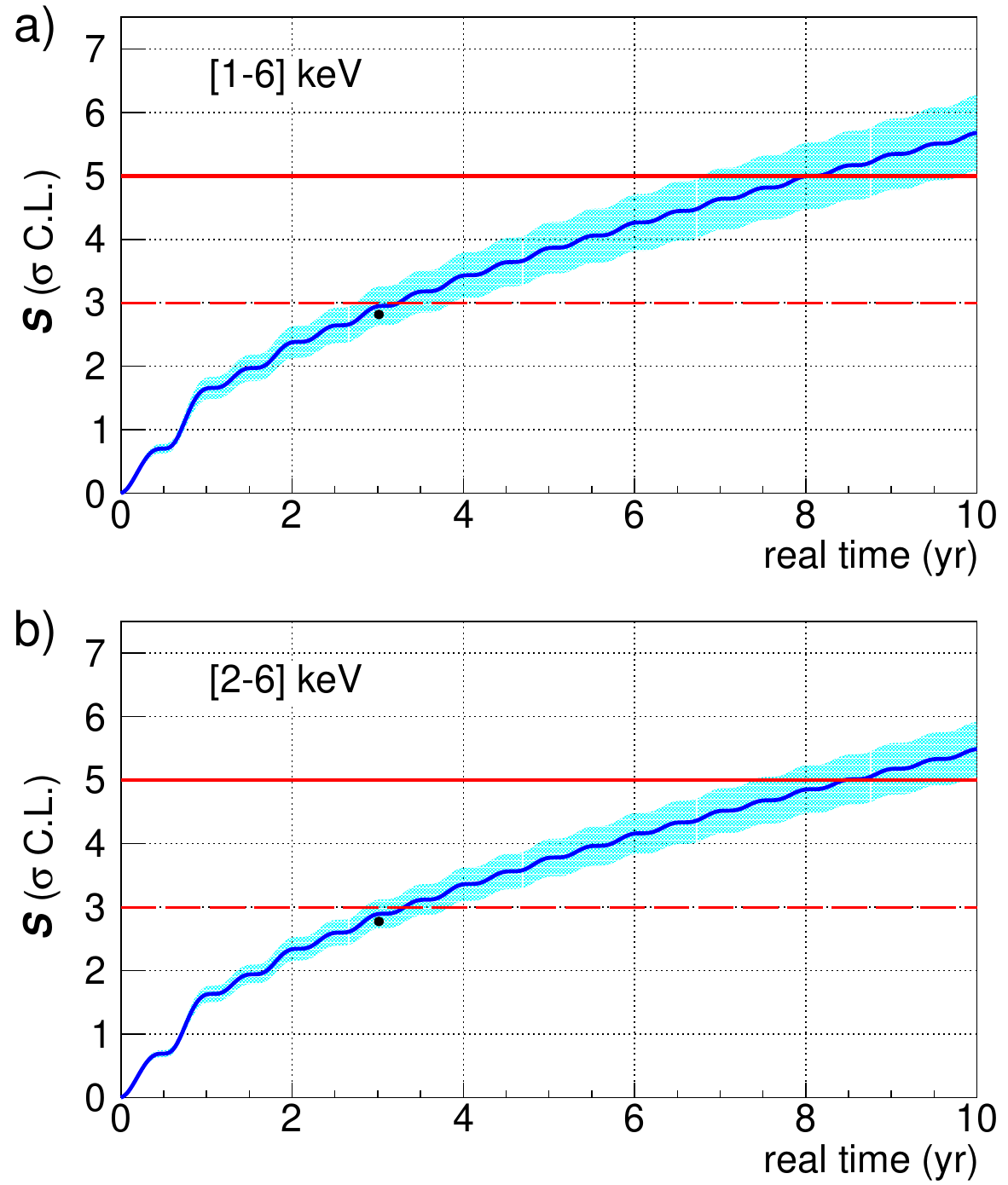} 
    \caption{
    \textbf{Evolution of the \ANAIS sensitivity to the \DL signal over time and sensitivity obtained in this work.}
    Dark blue line: the \ANAIS sensitivity to the \DL signal is represented in  $\sigma$ conficence level (C.L.) units as a function of real time in the [1--6]~keV (a) and [2--6]~keV (b) energy regions. The black dot is the sensitivity measured experimentally in this work. The cyan bands represent the 68\% C.L. \DL uncertainty. Red dashed and red solid lines correspond to reference values of 3$\sigma$ and 5$\sigma$, respectively.\label{fig:sens}}
   
\end{figure}

We assess our sensitivity to the DAMA/LIBRA signal as the ratio $S_m^{\text{DAMA}}/\sigma(S_m)$, 
which directly gives in $\sigma$ units the C.L. at which we can test the DAMA/LIBRA result.
The standard deviations for the modulation amplitude obtained in the best fit, 
\linebreak 
$\sigma(S_m)=3.7\,$\,counts\,keV$^{-1}$\,ton$^{-1}$\,d$^{-1}$ for both [1--6]\,keV and [2--6]\,keV, 
correspond to a sensitivity of (2.8$\pm$0.3)$\sigma$ in [1--6]\,keV and (2.8$\pm$0.2)$\sigma$ in [2--6]\,keV, where the uncertainty corresponds to the 68\% C.L. DAMA/LIBRA uncertainty.
\\
Fig.~\ref{fig:sens} displays in dark blue lines the \ANAIS sensitivity projections 
following Ref.~\cite{Coarasa:2018qzs}, conveniently updated to the effective 
exposure and detection efficiency presented in this work.
Cyan bands take into account the 68\% uncertainty in $S_m^{DAMA}$. 
The black dot is the sensitivity derived from the results presented here, 
in good concordance with our estimates.
These results support our 
expectation of achieving a 5$\sigma$ sensitivity by 2025.
%


\subsection*{Investigating the impact of the hypothesis of different quenching factors among detectors}

\begin{figure*}
    \centering
    \includegraphics[width=0.85\textwidth]{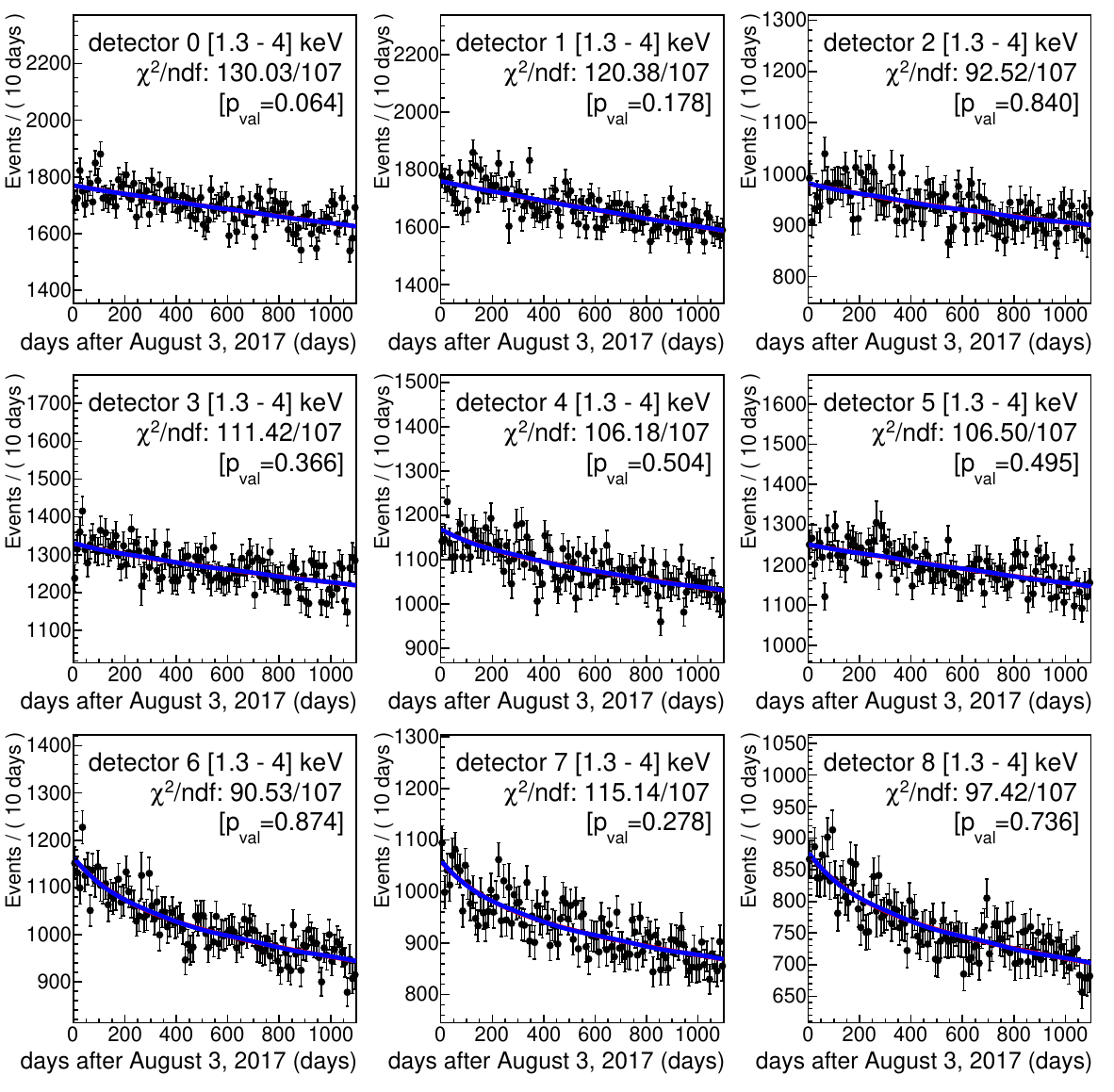} 
    \caption{
    \textbf{Fit results for data from the nine \ANAIS modules in the [1.3--4]\,keV energy range, under both the modulation and null hypotheses}. Each panel corresponds to one of the nine detectors of ANAIS-112. The error bars on the data points represent the standard deviation of the observed rate of events combined with the efficiency uncertainty. The blue line shows the result of the modulation hypothesis fit, while the red line represents the result of the null hypothesis, although it is generally masked by the blue line and not visible. Each panel also displays the $\chi^2$ divided by the degrees of freedom (NDF) of the fit for each detector, along with the corresponding p-value. The global results of the fit are: for the null hypothesis, $\chi^2$/NDF = 969.61/972, corresponding to a p-value~=~0.516, and for the modulation hypothesis, $\chi^2$/NDF~=~969.18/971, corresponding to a p-value~=~0.510. The best-fit modulation amplitude in the latter case is $S_m = (3.3 \pm 5.0)$ (counts keV$^{-1}$ ton$^{-1}$ day$^{-1}$).  }
    \label{fig:rateEvol1p34}
\end{figure*}

The importance of quenching factors has already been emphasized in comparing data 
from experiments with scintillators searching for WIMPs through their elastic interaction with atomic nuclei. Because of this, a model-independent testing of the \DL result requires using the same target material. Additionally, it is necessary to calibrate the detectors in nuclear-recoil energies, as far as using the conventional electron equivalent energy calibration cannot guarantee a fair comparison of the same energy regions in the case those quenching factors vary significantly for different NaI(Tl) detectors, for example, due to different concentrations of Tl or the presence of impurities or defects. 
\\
In recent decades, the community working with NaI(Tl) has made significant efforts to shed light on this issue. 
ANAIS and COSINE have conducted a joint study on quenching factors in crystals produced by Alpha Spectra, used by both experiments~\cite{Cintas:2024pdu}. 
Small crystals from the same supplier but with different powder quality were measured using a monochromatic neutron source at TUNL,
North Carolina (US). 
The results were consistent for all measured crystals. The study also highlighted the importance of properly considering 
the well-known non-linearity in the NaI(Tl) response, as it can distort the results at very low energies. 
The results of this work yield constant $Q_{\text{Na}}$ values of 0.210$\pm$0.003 or slightly decreasing with decreasing energy 
down to a value of $\sim$0.15 for recoil energies $E_{\text{Na}}$=10\,keV, depending on the calibration method. Despite the differences observed in the various measurements of $Q_{\text{Na}}$, 
there is a general consensus towards $Q_{\text{Na}}$  values on the order of 0.2, which decrease as energy decreases 
below $E_{\text{Na}}$=30\,keV to values around 0.10--0.15. 
In this regard, it is also interesting to mention the preliminary results obtained in Ref.~\cite{Bharadwaj:2023aoz} 
with 5 crystals with variable Tl dopant levels ranging from 0.1, 0.3, 0.5, 0.7 to 0.9\%, 
which point to values of $Q_{\text{Na}}$ in the range of 0.2 for all of them.
Few measurements deviate from this trend, 
such as Ref.~\cite{Spooner:1994ca} and 
notably Ref.~\cite{Bernabei:1996vj}, 
obtained on crystals from \DL 
through a $^{252}$Cf neutron calibration. Assuming energy-independent quenching factors, 
results were consistent with $Q_{\text{Na}}^{\text{DAMA}}$=0.3. 
\\
Concerning $Q_{\text{I}}$, the situation is similar, 
although due to its low value, measurements are more challenging and experimental results are scarce. 
The ANAIS and COSINE joint work~\cite{Cintas:2024pdu} has obtained a value of 0.060$\pm$0.022. Available data~\cite{Joo:2017aws, Xu:2015wha, Collar:2013gu} also point to $Q_{\text{I}}$ on the order of 0.06. Similarly, in this case, the value obtained by DAMA for the constant quenching factor hypothesis~\cite{Bernabei:1996vj} is higher ($Q_{\text{I}}^{\text{DAMA}}$=0.09).
\\
In conclusion, the possibility that the quenching factor of DAMA crystals differs from those observed in recent measurements remains open.
Taking into consideration this scenario for the comparison between \ANAIS and DAMA\slash LIBRA, 
if we aim to compare the same energy region in terms of  nuclear recoils, the \DL region from [2--6]\,keV 
would correspond to the \ANAIS region from [1.3--4]\,keV for constant $Q_{\text{Na}}$=0.2 and $Q_{\text{I}}$=0.06~\cite{Cintas:2024pdu}. 
We have carried out that analysis, 
the results of which are depicted in Fig.~\ref{fig:rateEvol1p34}. 
Once again, a high p-value is obtained for the null hypothesis, 
while the best fit provides a modulation amplitude of $S_m=3.3\pm5.0$\,counts\,keV$^{-1}$\,ton$^{-1}$\,d$^{-1}$.
In this case, the sensitivity is 3$\sigma$ because, although the statistics is reduced as a consequence of the reduction of the integration window, the signal-to-background ratio increases correspondingly.
%
\section*{Methods}
\label{sec:meth}
\subsection*{Energy calibration}
\label{ssec:ecal}

Energy calibration is carried out in two different ranges: high energy (HE) and low energy (LE). For both regimes, we have updated our calibration procedure with respect to~\cite{Amare:2018sxx}. In the case of HE, background measurements are used; while for LE, periodic calibrations are performed with a \Cd source which allows correction for possible gain drifts. Finally, the ROI is recalibrated using two lines from the background corresponding to \Na (0.87~keV) and \K (3.20~keV).
\subsubsection*{High energy calibration}
\label{sssec:hecal}

The digitization scale is optimized for the study of the low energy events, so the events above $\sim$500~keV are out of the digitizer dynamic range and pulse area energy estimator, $S_{sum}$, saturates because the digitized pulses (negative) are truncated at the minimum voltage (-1~V). For this reason, the \ANAIS DAQ system~\cite{Amare:2018sxx} incorporates a second signal line conveniently attenuated to retain information on the energy released by every high energy event through the use of charge-to-digital converters (QDC).

\begin{figure}
    \centering
    \includegraphics[width=0.48\textwidth]{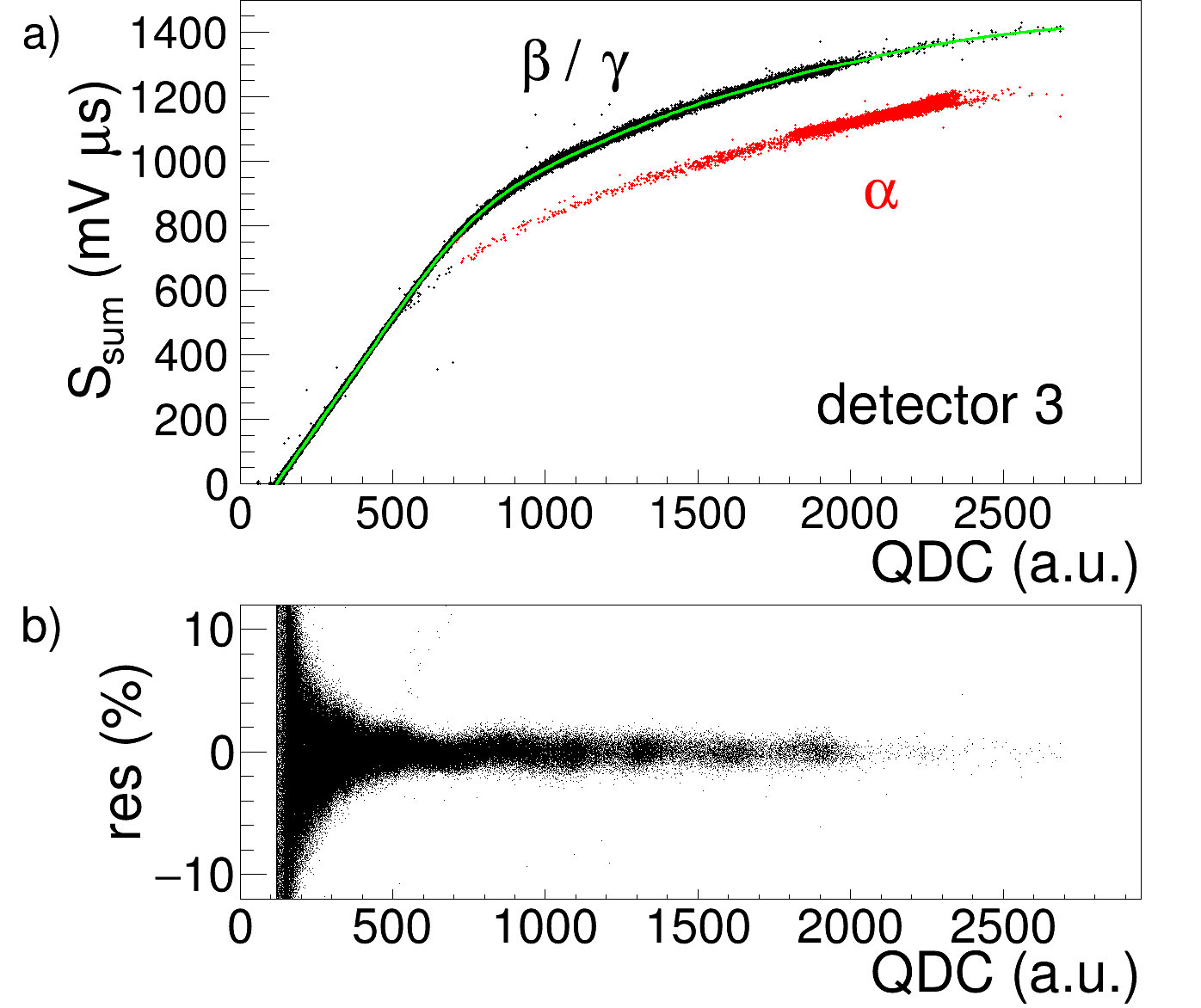} 
    \caption{\textbf{An example of the procedure for obtaining the high-energy estimator.} Dots in panel a) are the total pulse area for detector D3 during the two first weeks of data taking as a function of the QDC readout. The $\alpha$ population (red dots) is clearly separated from the $\beta/\gamma$ one (black dots). The green line is the result of a fit to a 12$^{\text{th}}$-order Chebyshev polynomial. The panel b) shows the residuals of the $\beta/\gamma$ population fit to the green line.}
    \label{fig:HEcalLinear}
\end{figure}

As an example, Fig.~\ref{fig:HEcalLinear} represents the pulse area versus the corresponding QDC value for detector D3 during the two first weeks of data taking. It can be observed that the pulse area parameter is clearly saturated above QDC$\simeq$700. In order to estimate the energy of events above $\sim$100~keV, the linearization of the $S_{sum}$ response was previously performed using a modified logistic function~\cite{Amare:2018sxx}, but the deviation of high-energy events from the fit reached up to 4\%. Therefore, we have updated the high-energy linearization function (green line) by combining a first-degree polynomial (QDC$<250$), a 12$^{\text{th}}$-order Chebyshev polynomial (approximately up to 90\% of the QDC saturation value), and a second-degree polynomial, successfully reducing the high-energy residuals below 2\% (as can be seen in the top panel).
%

This double readout system also allows to discriminate $\alpha$ events (shown in red in Fig.~\ref{fig:HEcalLinear}) from $\beta/\gamma$ events (depicted in black). For high energy events the digitized pulses are saturated and, as $\alpha$ events are faster than $\beta/\gamma$, the integral of the pulse in the microsecond window is smaller for the same QDC value.

Since there are no external sources of high energy available for calibrating the \ANAIS high energy regime, calibration of events above $\sim$100~keV is conducted independently for each background run using several easily identifiable peaks present in the background data. For every detector and run,
which in average lasts for two weeks, the number of peaks used for calibration ranges from 6 to 7, depending on their presence in the background spectrum. Among them are: 238.6~keV ($^{212}$Pb), 295.2~keV ($^{214}$Pb), 351.9~keV ($^{214}$Pb), 609.3~keV ($^{214}$Bi), 1120.3~keV ($^{214}$Bi), 1460.8~keV ($^{40}$K) and 1764.5~keV ($^{214}$Bi). Each peak is fitted to Gaussian lineshapes. Eventually, the calibration is performed via a linear regression between the nominal energies of the peaks and the Gaussian means using a second-degree polynomial.

\begin{figure}
    \centering
    \includegraphics[width=0.48\textwidth]{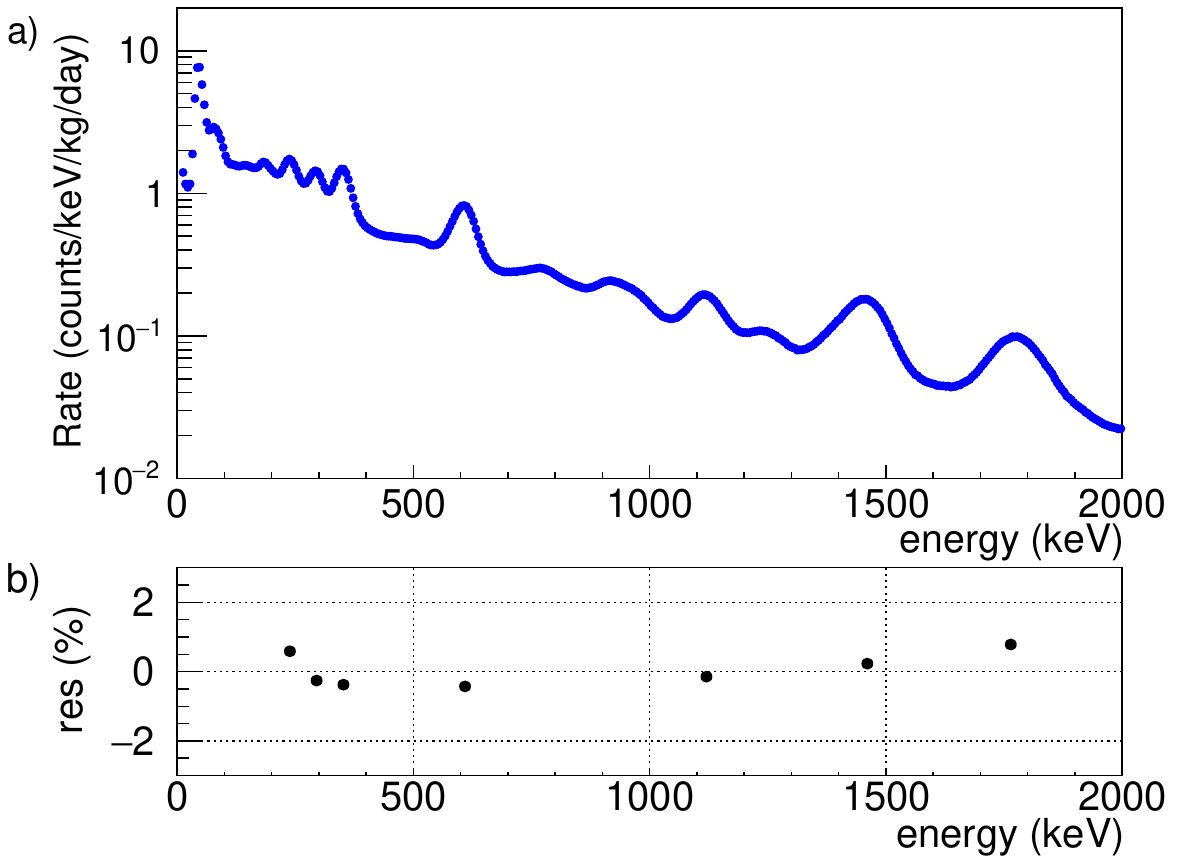} 
    \caption{\textbf{Total high energy anticoincidence spectrum measured along the first three years of \ANAIS operation.} Panel a) shows the spectrum and panel b) shows the residuals for the positions of the main peaks identified in the background.}
    \label{fig:HEcalSpc}
\end{figure}

Fig.~\ref{fig:HEcalSpc} shows the high-energy calibrated background spectrum for single-hit events adding all the \ANAIS detectors over the first three years of operation. Panel b) shows the residuals ((fit -- nominal)/nominal) for the positions of the main peaks identified in the background, being all of them below 1\%.
%
\subsubsection*{Low energy calibration}
\label{sssec:lecal}

The \ANAIS modules feature a Mylar window in the lateral face, allowing for the use of external gamma sources to perform low-energy calibration. Every two weeks, \Cd sources are introduced from outside the shielding via multiple flexible wires, enabling simultaneous calibration of all nine modules. \Cd decays by electron capture (EC) emitting a 88.0~keV $\gamma$. K$_\alpha$ and K$_\beta$ Ag X-rays are also emitted with average energies of 22.1 and 25.1~keV, respectively. In addition, the source plastic housing contains a certain amount of bromine, which under \Cd irradiation produces a new calibration line in correspondence with the K-shell Br X-rays. For the Br line, we take as nominal energy the average of the K$_\alpha$ and K$_\beta$ X-rays, resulting in 12.1~keV. The 12.1 and 88.0~keV lines are fitted to single Gaussian lineshapes added to a first-degree polynomial, while the 22.1+25.1~keV lines are fitted to two Gaussian lineshapes with the same standard deviation added to a first-degree polynomial. Then, a linear regression on the expected energies against the positions of the peaks for every detector is performed using a linear function, and the recalibration of the low energy events (below $\sim$100~keV) is carried out.

In order to increase the reliability of the energy calibration in the ROI, and trying to reduce possible non-linearity effects, we can also profit from two known lines present in the background, which are actually either in the ROI or very close to the threshold. These lines correspond to an internal contamination of \K in the bulk and the presence of \Na as a result of cosmogenic activation. These isotopes may decay via EC, with the emission of a $\gamma$ from the daughter nucleus de-excitation. The atomic de-excitation energy (0.87~keV for \Na and 3.2~keV for \K for K-shell EC, which has the largest probability) is fully contained in the crystal where the decay occurs, while the high energy $\gamma$ (1274.5 and 1460.8~keV, respectively) can escape and hit another detector, thus producing a coincidence event. 

The \NaK low-energy peaks are excellent for low energy calibration, but their low rate and the low efficiency for the detection of the coincidence to select them accurately requires the accumulation of background data over long periods to observe them properly. It is also worth noting that the \Na peak is below the analysis threshold, and despite efficiently triggering at the photoelectron level each PMT, the requirement of coincidence between the two PMTs within the 200~ns window results in a non-negligible decrease in trigger efficiency below 1~keV. This efficiency was estimated through a Monte Carlo simulation in~\cite{Amare:2018sxx} and has been used to correct the nominal energy of the \Na peak from 0.87~keV to 0.90~keV. Thus, we have accumulated low-energy coincident events of \Na and \K over the first five years of measurement for each detector, and each peak has been fitted to a Gaussian lineshape added to a first-degree polynomial to estimate its mean value. Eventually, the ROI calibration has been conducted via linear regression between the Gaussian means and the nominal energies (0.90 and 3.2~keV, respectively) using a linear function.

\begin{figure*}
    \centering
    \includegraphics[width=\textwidth]{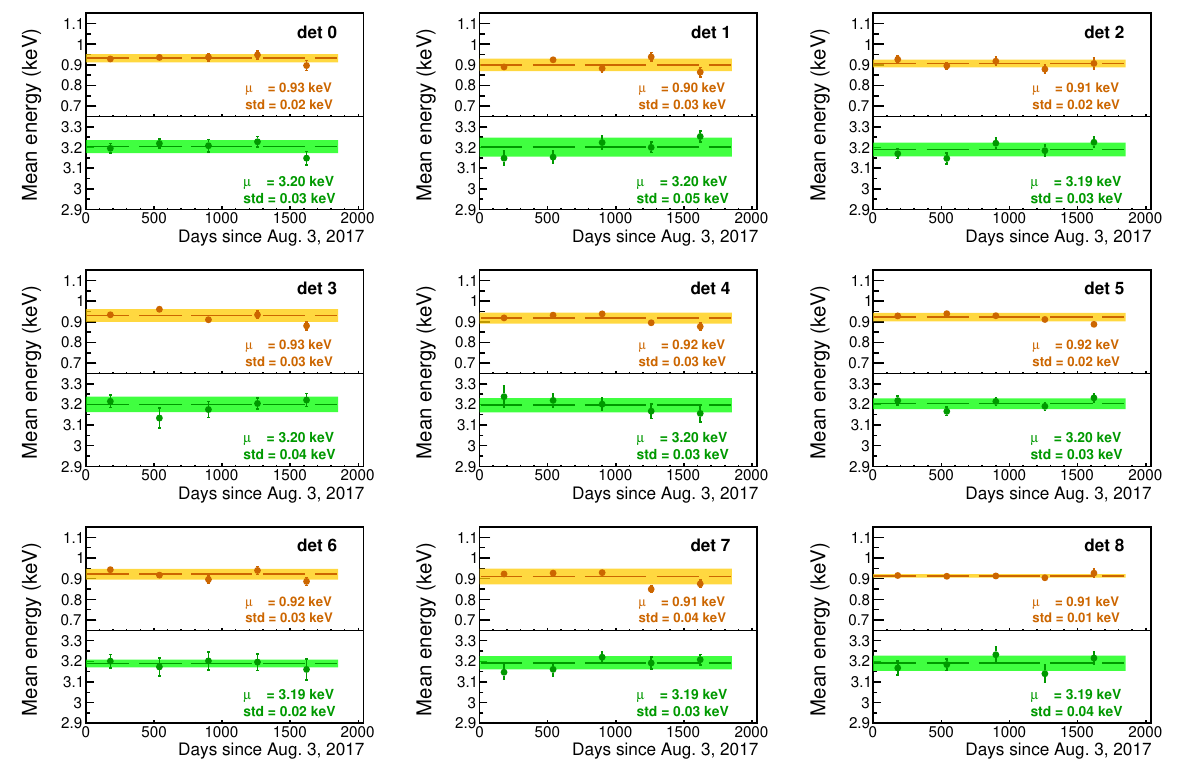} 
    \caption{\textbf{Stability over time of the calibration peaks in the region of interest for the nine \ANAIS detectors.} Upper panels, in orange: Evolution of the mean energy of the fitted 0.90~keV peak from \Na along the first five years of data taking for the nine \ANAIS modules. Lower panels, in green: the same, but for the 3.2~keV peak from $^{40}$K. Error bars represent the standard deviation of the mean value. The dashed lines are the mean values of the peak positions in each detector, and the shaded regions represent the standard deviations of the peak positions. The mean value and the standard deviation for each module are also shown in the panels.}
    \label{fig:NaKStab}
\end{figure*}

Fig.~\ref{fig:NaKStab} shows the evolution of the mean energy of the fitted 0.90 and 3.2~keV peaks from \Na (in orange) and \K (in green), respectively, along the first five years of data taking for the nine \ANAIS modules using this calibration strategy. It can be observed that the energy scale over time is stable within the ROI in all detectors.

%
\subsection*{Event selection}
\label{ssec:evsel}

The trigger rate in the ROI is dominated by non-bulk scintillation events. For this reason, the development of robust protocols for the selection of events corresponding to bulk scintillation in sodium iodide is mandatory. Initially, the selection criteria applied in \ANAIS were based on standard cuts on a few parameters~\cite{Amare:2018sxx}, and even though they demonstrated effectiveness above 2~keV, they showed weaknesses in the region from 1 to 2~keV. In order to improve the rejection of noise events between 1 and 2~keV, a machine-learning technique based on a Boosted Decision Tree (BDT) has been implemented. The detailed description of the BDT performance in \ANAIS with the old low-energy calibration can be found in~\cite{Coarasa:2022zak}. Because of the implementation of the new low-energy calibration, the BDT filtering method requires a new training procedure using the updated populations. As training populations for BDT, we combine the following: as signal events, scintillation events ranging from [1--2]~keV inside the crystal bulk produced by neutron interactions from dedicated neutron calibrations with a $^{252}$Cf source located outside the \ANAIS shielding, which are predominantly associated with elastic nuclear recoils in that energy region~\cite{Pardo:2023rrd}; and as noise events, those coming from a blank module similar to the \ANAIS modules, but without NaI(Tl) crystal. This choice of the training populations is robust, as it entirely excludes background events, and uses bulk events as signal.
The fact that we do not have pure scintillation populations can slightly bias the training, although the major effect is underestimating the cut efficiency.
The training process results in a newly constructed parameter named BDT, which combining the information of 15 discriminating parameters maximizes the separation between the signal and noise populations used in the training. For the event selection, we define an energy-dependent BDT cut, retaining only those events that exceed it. This selection criterion has been fine-tuned for each detector and energy bin (see \cite{Coarasa:2022zak} for details). The corresponding efficiency is estimated individually for each detector by using $^{252}$Cf neutron calibration events. The ratio of events that pass the signal selection to the total events determines the acceptance efficiency which, when multiplied by the trigger efficiency, constitutes the total detection efficiency. Total efficiency for event detection in all the \ANAIS modules as a function of energy is shown in Fig.~\ref{fig:eff}. The acceptance efficiency derived from the BDT cut notably exceeds (around 30\% in [1--2]~keV) that of the previous \ANAIS filtering procedure. Moreover, the BDT method significantly diminishes the background level below 2~keV for all detectors compared to that obtained using the previous \ANAIS protocols. In particular, the integrated rate from 1 to 2~keV is 5.39$\pm$0.04 and 4.40$\pm$0.03\,counts\,keV$^{-1}$\,kg$^{-1}$\,d$^{-1}$ for the \ANAIS filtering procedure and the BDT method, respectively, representing an 18\% reduction in background.
\subsection*{Trigger rate cut}
\label{ssec:rateCut}

\begin{figure*}
    \centering
    \includegraphics[width=\textwidth]{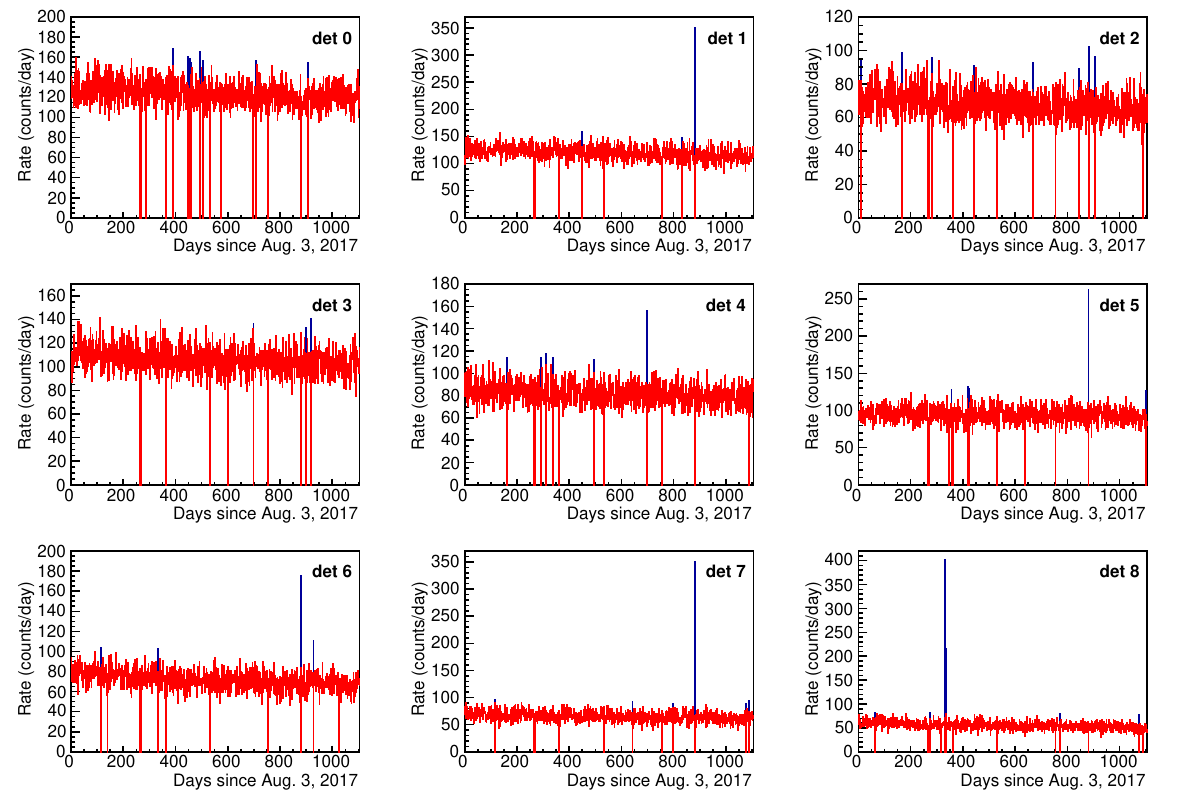} 
    \caption{\textbf{Evolution over time of the trigger rate for each \ANAIS detector.} In blue: trigger rate (calculated in 1-day time bins) for events filtered with the Boosted Decision Tree (BDT)  algorithm below 3~keV during the first three years of \ANAIS data taking. In red: the same, but after applying the cut in trigger rate described in the text. Note that usually the blue line is masked by the red one.}
    \label{fig:triggerRate}
\end{figure*}

Radioactive backgrounds and dark matter are expected to produce a constant rate of events in the detector when considering short time intervals. High-trigger rate periods may be caused by, for example, electrical or mechanical disturbances. In this scenario, high-rate periods, statistically inconsistent with the average detection rate of the experiment can be safely discarded. In order to do so, we evaluated the daily rate of low-energy single-hit events (below 3~keV) passing the BDT cut for each detector (blue line in Fig.~\ref{fig:triggerRate}). All bins exceeding three standard deviations from the detector's annually averaged rate are removed. The red line in Fig.~\ref{fig:triggerRate} shows the event rate surviving the cut. The fraction of rejected live time after applying this filtering varies for each detector but, in any case, it is less than 1\%.
%

\section*{Data availability}

The data used for this manuscript are available at the website of the ORIGINS Excellence Cluster: \url{https://www.origins-cluster.de/odsl/dark-matter-data-center/available-datasets/anais}.


\section*{Code availability}

The scripts for reproducing all the figures and results are available at the website of the ORIGINS Excellence Cluster: \url{https://www.origins-cluster.de/odsl/dark-matter-data-center/available-datasets/anais}.


\section*{References}
\bibliography{biblio}
\section*{Acknowledgements}

This work has been financially supported by MCIN/AEI/10.13039/501100011033 under grant PID2022-138357NB-C21 and PID2019-104374GB-I00, the Consolider-Ingenio 2010 Programme under grants MultiDark CSD2009-00064 and CPAN CSD2007-00042, the LSC Consortium, the Gobierno de Arag\'on and the European Social Fund (Group in Nuclear and Astroparticle Physics) and funds from European Union NextGenerationEU/PRTR (Planes complementarios, Programa de Astrof\'{\i}sica y F\'{\i}sica de Altas Energ\'{\i}as). Authors would like to acknowledge the use of Servicio General de Apoyo a la Investigaci\'on-SAI, Universidad de Zaragoza and technical support from LSC and GIFNA staff.


\section*{Author contributions statement}

E.G., A.O.d.S. and M.L.S. contributed to the design of the experiment; J.A., S.C., M.M., M.\'A.O., A.O.d.S., A.S., M.L.S. and P.V. contributed to the setting-up and commissioning of the experiment; J.A., J.Ap., S.C., D.C., I.C., M.M., M.Á.O., Y.O., A.O.d.S., T.P. and M.L.S. performed calibration and maintenance of the experiment; M.\'A.O. and M.M. developed the DAQ software tools; I.C., M.M., M.Á.O. and M.L.S. developed analysis tools; I.C., E.G., M.M., J.P. and M.L.S. analyzed the data; S.C. and P.V. developed background simulation codes; E.G., A.O.d.S. and J.P. performed radiopurity measurements; I.C., E.G., M.M. and J.P. contributed to sensitivity estimates. The manuscript and plots were produced by I.C. and M.M. All authors have read and agreed to the published version of the manuscript. Authors are listed alphabetically by their last names.

\section*{Competing Interests}

The authors declare no competing interests.


\end{document}